\documentclass[twocolumn, astrosymb]{aastex631}

\newcommand{\sn}{SN~2016geu}

\shorttitle{Sodium towards SN~2016geu}
\shortauthors{Gall et al.}
\graphicspath{{./}{Figures/}}

\begin{document}

\title{Origin of the strong sodium absorption of the lensed supernova 2016geu at z=0.4}

\author[0000-0002-8526-3963]{Christa Gall}
\affiliation{DARK, Niels Bohr Institute, University of Copenhagen, Jagtvej 155, DK-2200 Copenhagen N, Denmark}

\author[0000-0002-4571-2306]{Jens Hjorth}
\affiliation{DARK, Niels Bohr Institute, University of Copenhagen, Jagtvej 155, DK-2200 Copenhagen N, Denmark}

\author[0000-0001-8415-7547]{Lise Christensen}
\affiliation{Cosmic Dawn Center, Niels Bohr Institute, University of Copenhagen, Jagtvej 155, DK-2200 Copenhagen N, Denmark}

\author[0000-0001-9695-8472]{Luca Izzo}
\affiliation{INAF, Osservatorio Astronomico di Capodimonte, Salita Moiariello 16, I-80131 Napoli, Italy}
\affiliation{DARK, Niels Bohr Institute, University of Copenhagen, Jagtvej 155, DK-2200 Copenhagen N, Denmark}

\author[0000-0001-6876-8284]{Paolo A. Mazzali}
\affiliation{Astrophysics Research Institute, Liverpool John Moores University, ic2, 146 Brownlow Hill, Liverpool, L3 5RF, United Kingdom}
\affiliation{Max-Planck Institut für Astrophysik, Karl-Schwarzschild-Str. 1, 85741, Garching, Germany}

\author[0000-0003-2734-0796]{Mark M. Phillips}
\affiliation{Carnegie Observatories, Las Campanas Observatory, Casilla 601, La Serena, Chile}

\author[0000-0002-4338-6586]{Peter Hoeflich}
\affiliation{Florida State University, Tallahassee, Florida, USA}

\author[0000-0002-4269-7999]{Charlotte Angus}
\affiliation{Astrophysics Research Centre, School of Mathematics and Physics, Queen’s University Belfast, Belfast BT7 1NN, UK}
\affiliation{DARK, Niels Bohr Institute, University of Copenhagen, Jagtvej 155, DK-2200 Copenhagen N, Denmark}

\author[0000-0001-7666-1874]{Cecilie Cold}
\affiliation{DARK, Niels Bohr Institute, University of Copenhagen, Jagtvej 155, DK-2200 Copenhagen N, Denmark}

\author[0000-0001-9058-3892]{Jonatan Selsing}
\affiliation{DARK, Niels Bohr Institute, University of Copenhagen, Jagtvej 155, DK-2200 Copenhagen N, Denmark}







\begin{abstract}

The origin of strong sodium absorption, which has been observed for a few nearby Type Ia supernovae (SNe~Ia), remains elusive. 
Here we analyse two high-signal-to-noise, intermediate-resolution VLT/X-shooter spectra at epochs $+$18 and $+$27 days past peak brightness of the strongly lensed and multiply-imaged Type Ia \sn\, which exploded at a redshift of $z=0.4$.
We show that \sn\ exhibits very strong, multiple Na~{\small{I}} and Ca~{\small{II}} absorption lines with a large total Na~{\small{I}}~D restframe equivalent width of 5.2 $\pm$ 0.2 \AA, among the highest ever detected for a SN Ia and similar to only a handful of nearby SNe~Ia with 
extraordinary large Na~{\small{I}}~D EWs.  
The absorption system is time-invariant and extends over a large velocity span $\sim$ 250 km s$^{-1}$. 
The majority of the absorption is blueshifted relative to the strongest component, while there are both blueshifted and redshifted components relative to the systemic redshift of the galaxy. 
The column density ratios and widths of the absorption lines indicate that the absorption likely arises from a combination of interstellar dusty molecular clouds and circumgalactic in- and outflowing material, 
rather than circumstellar matter around the SN. 
\end{abstract}


\section{Introduction} 
\label{sec:intro}
Type Ia supernovae are amongst the most powerful explosions in the universe. Their progenitors consist of a close binary stellar system in which a carbon-oxygen degenerate white dwarf undergoes a thermonuclear runaway \citep[e.g.,][]{1960ApJ...132..565H,2000ARA&A..38..191H,2011cetd.book.....H}. While the classical scenarios include the double-degenerate \citep[DD, ][]{1984ApJS...54..335I,1984ApJ...277..355W} model and the single-degenerate model \citep[SD, ][]{1973ApJ...186.1007W,1982ApJ...253..798N}, 
a wide variety of progenitor systems and mechanisms triggering the terminal explosion have been proposed \citep[for a review see][and references theirein]{2018PhR...736....1L}.

Complex Na~{\small{I}}~D absorption systems with multiple components have been observed in several SNe~Ia in the nearby Universe \citep{1989A&A...215...21D, 2009ApJ...702.1157S, 2009ApJ...693..207B, 2011Sci...333..856S, 2012ApJ...752..101F, 2013ApJ...779...38P, 2013MNRAS.436..222M, 2015ApJ...799..197R}. 
Blueshifted components have traditionally been attributed to 
circumstellar material that has been ejected during the 
pre-supernova phase \citep{2011Sci...333..856S, 2013ApJ...779...38P,
2013MNRAS.436..222M, 2021MNRAS.507.4367C}. 
In some of these SNe~Ia with blueshifted Na~{\small{I}}~D components, e.g., SN~2006X \citep{2007Sci...317..924P}, SN~2007le \citep{2009ApJ...702.1157S} or iPTF11kx \citep{2012Sci...337..942D}, 
the absorption strength, or equivalent width (EW), was found to
vary 
over several days around peak brightness of the supernova. The time variability has been explained as an effect of ionisation and subsequent recombination of Na~{\small{I}}~D in shells or clumps of outflowing circumstellar material \citep{2007Sci...317..924P}. The latter is the relic of pre-supernova evolution
\citep{2007Sci...317..924P, 2011A&A...530A..63P, 2012Sci...337..942D, 2013ApJ...772....1R, 2013MNRAS.431.1541S}.  
In other systems, e.g., SN 1986G \citep{1989A&A...215...21D} and SN 2014C \citep{2015ApJ...799..197R}, strong Na~{\small{I}}~D absorption has been attributed to interstellar material.

Aided by strong magnification due to gravitational lensing, the 
multiply-imaged \sn\ at $z=0.4$ can shed new light on the origin of strong Na~{\small{I}}~D absorption.
\sn\, was discovered by
the intermediate Palomar Transient Factory (iPTF) on 5 September, 
2016 in the SDSS galaxy J210415.89-062024.7 and was classified as a 
normal SN Ia \citep{2017Sci...356..291G, 2018MNRAS.473.4257C, 2021MNRAS.502..510J}. 
A thorough spectroscopic analysis \citep{2018MNRAS.473.4257C} 
revealed that \sn\ belongs to groups of both high-velocity and 
high-velocity-gradient \citep{2005ApJ...623.1011B} and 
core-normal SNe~Ia \citep{2006PASP..118..560B}. 
It appeared $\sim 70$ times brighter than an unlensed SNe~Ia at the 
same redshift due to gravitational lensing (and possibly microlensing) 
magnification \citep{2017Sci...356..291G, 2017ApJ...835L..25M, 
2020MNRAS.491.2639D}. 
We obtained 
two optical to near-infrared medium-resolution VLT/X-shooter spectra \citep{2018MNRAS.473.4257C,2021MNRAS.502..510J} past peak brightness.
The large magnification
led to an unprecedented signal-to-noise ratio of the 
medium-resolution spectra of a $z=0.4$ supernova. 

Here we present a detailed analysis of the Na~{\small{I}}~D and Ca H\&K absorption line complex of \sn. 
The Na~{\small{I}}~D complex was previously studied by \citet{2021MNRAS.502..510J}.
The paper is organised as follows: the data and the analysis of the 
Na~{\small{I}}~D and Ca H\&K absorption complex are 
described in Section \ref{sec:data} and 
Section~\ref{sec:NaCacomplex}, respectively. Other 
absorption and emission lines are analysed and 
discussed in Section~\ref{sec:origin}. We discuss 
the origin of the individual Na~{\small{I}}~D and 
Ca H\&K absorption components in Section~\ref{s:discussion}
and conclude in Section~\ref{sec:conclusion}.
\section{Data} 
\label{sec:data}
%
The optical to near-infrared medium-resolution 
spectra of \sn\ \citep{2018MNRAS.473.4257C} were obtained with the 
X-shooter echelle spectrograph \citep{2006SPIE.6269E..98D, 2011A&A...536A.105V} 
mounted at the Cassegrain focus of the {\it Kueyen} unit of the Very
Large Telescope (VLT) at the European Southern Observatory (ESO) on 
Cerro Paranal, Chile. 
They were obtained at two epochs, 2016 Oct 18.02 UT (57679.02 Modified 
Julian Date (MJD)) and 2016 Oct 30.03 UT (57691.03 MJD) which 
correspond to restframe epochs of about $+$18.6 and $+$27.1 
days past peak brightness \citep{2017Sci...356..291G, 
2020MNRAS.491.2639D}, respectively. 
Both observations were performed at a position angle of 86 degree 
following the convention `North up - East left', 
with `ABBA' nodding between exposures along the 11$''$ slit. 

We used the ESO / {\tt ESOReflex} workflow versions 2.9.2 of the 
X-shooter pipeline \citep{2010SPIE.7737E..56M, 
2013A&A...559A..96F} to reduce the supernova and the (telluric) 
standard star spectra to two-dimensional bias-subtracted, flat-field
corrected, order rectified, wavelength and flux calibrated spectra. 
All wavelengths were calibrated in the vacuum frame. Throughout, line
transitions are discussed in the astrophysical common notation of 
air wavelength, with conversions between the two systems performed 
as given in \citet{1991ApJS...77..119M}. 
To obtain one-dimensional spectra, the two-dimensional spectra from 
the pipeline were optimally extracted using the full extraction 
profile for the spectral trace \citep{1986PASP...98..609H}. 
Furthermore, the spectra were slit-loss corrected and corrected for 
heliocentric velocities. Telluric corrections were applied using 
molecfit \citep{2015A&A...576A..77S, 2015A&A...576A..78K}. All 
calibrations and correction procedures were performed after the 
basic pipeline reduction using refined custom {\tt python} 
programs 
\citep{2019A&A...623A..92S}. The spectra were corrected for a 
Galactic extinction along the line of sight to the SN of $E(B-V)$ of 
0.31 $\pm$ 0.05 mag for $R_V$=3.1 \citep{2011ApJ...737..103S}. 
The spectra were lens galaxy subtracted using the lens model of \citet{2017Sci...356..291G} as described in \citet{2018MNRAS.473.4257C}. 
%
\section{The Sodium and Calcium Absorption Line Complex} 
\label{sec:NaCacomplex}
%
\sn\ exhibits prominent sodium and calcium absorption line complexes. 
The sodium doublet, Na~{\small{I}}~D, 
corresponds to the fine structure splitting of the neutral sodium 
excited states observed in air at $\lambda\lambda$5889.95 (D$_2$) and 5895.92 (D$_1$) \AA\, respectively. The calcium doublet, Ca~{\small{II}}~H 
\& K, corresponds to the fine structure splitting of the singly 
ionized calcium excited states observed in air at $\lambda\lambda$3968.46 (K) and 3933.65 (H) \AA. We use a redshift of $z$ = 0.408788 
(Na~{\small{I}}~D component B) as the systemic redshift of the host galaxy and the reference redshift for all components. 

\subsection{Analysis of the Absorption Line Components} 
\label{sec:ANA}

We identify and estimate the column densities of Na~{\small{I}}~D and 
Ca~{\small{II}}~H \& K components using the {\tt python} package {\tt VoigtFit} \citep{2018arXiv180301187K}. The continuum normalization 
is done interactively using 2$^{nd}$ and 5$^{th}$ order Chebyshev 
polynomials for Na~{\small{I}}~D and Ca~{\small{II}} H \& K, 
respectively. Higher order polynomials are required since the 
absorption lines are on top of stronger, high velocity absorption 
features from SN~Ia ejecta elements. 
As per definition of {\tt VoigtFit}, the main parameters, e.g. the 
ratio of a transition's 
oscillator strengths, $f$-values, are fixed such that the relative 
velocity, Doppler broadening parameter, $b$, and column density
for different transitions of the same ionization state are identical.
This means that the fitted parameters such as the $b$-values, column 
densities and the relative velocities of Na~{\small{I}}~D$_1$ and 
Ca~{\small{II}}~H will be the same as those of Na~{\small{I}}~D$_2$ 
and Ca~{\small{II}}~K. 

The EW and full width at half maximum (FWHM) of 
individual components of Na~{\small{I}}~D and Ca~{\small{II}}~H \& K 
are derived from the Voigt profile fits to the absorption lines. 
The EWs are calculated as 
    $W_{\lambda} = \int^{\lambda_1}_{\lambda_2} (1 - \frac{F_{\lambda}}{F_{\mathrm{C}}}) \, d_{\lambda}$,
with $F_{\lambda}$ and $F_{\mathrm{C}}$ being the observed flux of the line
and the continuum flux within the wavelength interval [$\lambda_1$, 
$\lambda_2$], respectively. 
The FWHM of the Voigt profiles is estimated as the full width at 
half the peak intensity of the line. For other ions the FWHM and 
minimum absorption velocities are derived from either Gaussian
or Lorentzian profile fits to the absorption lines. 
All measured 
quantities of absorption lines, including those of Na~{\small{I}}~D 
and Ca~{\small{II}}~H \& K of the lens galaxy are 
summarized in 
Tables~\ref{tab:LINDAT1} and \ref{tab:LINDAT2}. 

\begin{deluxetable*}{cclcccccc}[htb!]
\label{tab:LINDAT1}
\tablecaption{
 {\bf Measured properties of individual absorption lines.} }
\linespread{1.1}\selectfont\centering
\tablewidth{0pt}
\tablehead{
\colhead{Component} &
\colhead{Ion}  &
\colhead{redshift{$^a$}}     &
\colhead{EW{$^b$}}   &
\colhead{D$_{2}$/D$_{1}$}     &
\colhead{FWHM}    &
\colhead{log($N$)}   &
\colhead{$b$}   &
\colhead{$v_{\rm{rel}}${$^c$}}  \\
\nocolhead{} &  
\nocolhead{} &
\nocolhead{} &
\colhead{[\AA]}    &
\colhead{K / H}     &
\colhead{[km s$^{-1}$]}    &
\colhead{[cm$^{-2}$]}   &
\colhead{[km s$^{-1}$]}    &
\colhead{[km s$^{-1}$]}}
\startdata 
\multicolumn{9}{c}{\bf MJD 57679.0075}\\
\hline
\vspace{-0.2cm} & Na {\small{I}} D$_1$ & & 0.65(12) & & 47.3(1.4) & & & \\ \vspace{-0.2cm}
A & & 0.408407(3) & & 1.34(31) & & 13.02(0.03) & 19.98(1.08) & $-$81.0(0.6) \\ 
                & Na {\small{I}} D$_2$ & & 0.88(12) & & 52.6(1.3) & & & \\ 
\vspace{-0.2cm} & Na {\small{I}} D$_1$ & & 0.40(12)& & 45.4(2.6) & & & \\ \vspace{-0.2cm} 
B & & 0.408788(6)  & & 1.62(60) & & 12.71(0.03) & 19.95(2.04) & 0.0(0.0) \\ 
                & Na {\small{I}} D$_2$ & & 0.65(12) & & 48.0(2.6) & & & \\ 
\vspace{-0.2cm} & Na {\small{I}} D$_1$ & & 0.62(13) & & 84.3(36.6) & & &  \\ \vspace{-0.2cm} 
C1 & & 0.409287(86)&  & 1.7(43) & & 12.89(0.30) & 44.44(11.27)& 106.1(18.3) \\ 
                & Na {\small{I}} D$_2$ & & 1.07(15)& & 89.8(36.6)& & & \\ 
\vspace{-0.2cm} & Na {\small{I}} D$_1$ & & 0.44(13) & & 31.1(6.5) & & &  \\ \vspace{-0.2cm}
C2 & & 0.409378(15) & &1.2(48) &  & 15.68(1.93) & 3.04(2.6) & 125.6(3.2) \\ 
                & Na {\small{I}} D$_2$ & & 0.56(13) & & 33.8(6.3) & & & \\ 
\vspace{-0.2cm} & Na {\small{I}} D$_1$ & & 0.17(12)& & 30.9(7.4) & &  & \\ \vspace{-0.2cm}  
D & & 0.409595(17) & & 1.5(132) & & 12.31(0.20) & 8.67(8.25) & 171.7(3.6) \\
                & Na {\small{I}} D$_2$ & & 0.25(13)& & 31.3(7.5) & & & \\ 
\vspace{-0.2cm}  & Na {\small{I} D$_1$ }& & 0.82(11)& &57.8(5.4) & & & \\ \vspace{-0.2cm} 
 Lens  & &0.21628(22) & &1.6(26) & &12.91(0.04) & 96.10(4.25) & 2.47(2.47) \\ 
                & Na {\small{I}} D$_2$ & &1.36(11)& &59.9(5.5) & & & \\ 
%
\hline
\vspace{-0.2cm} & Ca {\small{II}} H & & 0.34(12)&  & 79.25(6.4) &  &  & \\ \vspace{-0.2cm} 
A  & &0.408412(15)& & 1.74(71) & & 12.94(0.05) & 39.35(5.92) & $-$80.0(3.2) \\ 
                & Ca {\small{II}} K & & 0.59(12)& & 83.23(6.4) & & & \\
\vspace{-0.2cm} & Ca {\small{II}} H & & 0.42(12)& & 74.89(19.6)& & &  \\ \vspace{-0.2cm}
C1  & & 0.409299(46) & & 1.6(54) & & 13.07(0.12) & 35.49(10.60) & 108.8(9.8) \\ 
                & Ca {\small{II}} K & & 0.67(12)& & 79.65(19.6)& & & \\ 
\vspace{-0.2cm} & Ca {\small{II}} H & & 0.14(12)&  & 54.15(30.6)& & &  \\ \vspace{-0.2cm} 
D  &  & 0.409582(72) & & 2.0(195) & & 12.58(0.33) & 22.17(16.71) & 169.1(15.3) \\ 
               & Ca {\small{II}} K & & 0.28(13) & & 54.71(30.6) & & & \\ 
Lens                   & Ca {\small{II}} H\&K & 0.2162(112) & 11.62(29)&$\cdots(\cdots)$&$\cdots(\cdots)$&$\cdots(\cdots)$&$\cdots(\cdots)$& $\cdots(\cdots)$\\ 
\hline
\multicolumn{9}{c}{\bf MJD 57691.0131}\\
\hline
\vspace{-0.2cm} & Na {\small{I}} D$_1$ & & 0.60(12) & & 52.3(2.6) & & & \\ \vspace{-0.2cm} 
A & & 0.408409(6) & & 1.5(36) & & 12.93(0.04) & 23.8(2.00) & $-$80.6(1.3) \\
                  & Na {\small{I}} D$_2$ & & 0.92(12)& & 57.2(2.6) & & & \\ 
\vspace{-0.2cm} & Na {\small{I}} D$_1$ &  & 0.35(12)& & 40.1(3.9) & & & \\ \vspace{-0.2cm} 
B  & & 0.408787(9) & & 1.6(64) & & 12.66(0.05) & 16.83(3.22) & $-$0.2(0.2) \\ 
                & Na {\small{I}} D$_2$ & & 0.57(12)& & 42.5(3.9) & & & \\ 
\vspace{-0.2cm} & Na {\small{I}} D$_1$ & & 0.25(13)& & 52.4(19.6) & & &  \\ \vspace{-0.2cm} 
C1 & & 0.40917(46) & & 1.8(106) & & 12.48(0.21) & 25.96(11.9) & 81.3(9.8)  \\ 
                & Na {\small{I}} D$_2$ & & 0.46(13)& & 54.3(19.6)& & & \\ 
\vspace{-0.2cm} & Na {\small{I}} D$_1$ & & 0.80(13)& & 39.6(4.7) & & & \\ \vspace{-0.2cm} 
C2 &  & 0.409368(11) & &1.2(26) & & 16.27(0.32) & 4.33(1.06) & 123.4(2.3)  \\ 
                & Na {\small{I}} D$_2$ & & 0.98(13)& & 45.1(4.7) & & & \\ 
\vspace{-0.2cm} & Na {\small{I}} D$_1$ & & 0.16(13)& & 27.9(5.1) & & & \\ \vspace{-0.2cm} 
D  &  & 0.409593(12) & & 1.4(142) & & 12.43(0.34) & 3.95(2.56) & 171.3(2.4) \\ 
                & Na {\small{I}} D$_2$ & & 0.21(13)& & 27.8(5.1) & & & \\
\vspace{-0.2cm}  & Na {\small{I}} D$_1$ & & 0.72(11)& & 67.9(8.4) & & & \\ \vspace{-0.2cm} 
Lens  & & 0.21628(28) & & 1.7(30) & & 12.87(0.06) & 87.32(10.17) & 9.37(4.2)  \\ 
                & Na {\small{I}} D$_2$ & & 1.22(11)& & 71.5(8.4) & & & \\ 
\hline
\vspace{-0.2cm}  & Ca {\small{II}} H & & 0.41(12) & & 79.73(8.9)  & & & \\ \vspace{-0.2cm}  
A  & & 0.408415(21) & & 1.76(60) & & 13.07(0.07) & 38.63(7.61) & $-$79.2(4.5) \\  
                & Ca {\small{II}} K & & 0.72(12)& & 84.31(8.9)  & & & \\ 
\vspace{-0.2cm}  & Ca {\small{II}} H & & 0.57(12)& & 139.92(14.1) & & & \\ \vspace{-0.2cm}   
C1  & & 0.409303(33) & & 1.74(42) & & 13.17(0.07) & 75.99(11.78) & 109.7(7.1) \\ 
                & Ca {\small{II}} K & & 0.99(12)& & 147.63(14.1) & & & \\ 
Lens                   & Ca {\small{II}} H\&K & 0.21621(77) & 14.92(28)&$\cdots(\cdots)$&$\cdots(\cdots)$&$\cdots(\cdots)$&$\cdots(\cdots)$&$\cdots(\cdots)$\\ 
\enddata
\tablecomments{
All parameters are obtained using {\tt VoigtFit} \citep{2018arXiv180301187K} and a custom {\tt IDL} program.
$^{a}${Errors are multiplied by 10$^6$.}
$^{b}${EWs are in restframe. Errors for the restframe EWs are multiplied by 10$^2$.}
$^{c}${All velocities are with respect to either $z$ = 0.408788, corresponding to component B of Na~{\small{I}}~D or $z$ = 0.2162 for the lens galaxy.}
}
\end{deluxetable*}
%

\begin{deluxetable*}{lcccccc}[htb!]
\label{tab:LINDAT2}
\tablecaption{
 {\bf Measured properties of absorption complexes.} }
\linespread{1.1}\selectfont\centering
\tablewidth{0pt}
\tablehead{
\colhead{Ion} &
\colhead{EW [\AA]{$^a$}}    &
\colhead{D$_{2}$/D$_{1}${$^a$}}     &
\colhead{EW [\AA]{$^a$}}   &
\colhead{D$_{2}$/D$_{1}${$^a$}}   &
\colhead{log($N$) [cm$^{-2}$] }   &
\colhead{log($N$) [cm$^{-2}$]}    \\
\colhead{Epoch}  &
\colhead{18.6 d}   &
\colhead{18.6 d}   &
\colhead{27.1 d}   &
\colhead{27.1 d}   &
\colhead{18.6 d}   &
\colhead{27.1 d}} 
\startdata 
\noalign{\smallskip}
\hline
\noalign{\smallskip}
Na {\small{I}} D  & 5.2(21) & & 5.06(26) & & 15.68(1.93) & 16.26(0.58) \\ 
\vspace{-0.2cm} Na {\small{I}} D$_1${$^b$} & 2.29(28)   &  &  2.16(28) & & &  \\ \vspace{-0.2cm} 
&  & 1.48(22) &  & 1.44(23) & &  \\ 
Na {\small{I}} D$_2${$^b$} & 3.40(29) &  & 3.13(28) & & &    \\ 
\hline
\noalign{\smallskip}
 &
 &
K / H     &
  &
K / H    &
 &
 \\
\noalign{\smallskip}
\hline
Ca {\small{II}} H\&K &  &    &  &   & 13.38(0.34) & 13.42(0.22) \\ 
\vspace{-0.2cm} Ca {\small{II}} H & 0.90(22)  & & 0.98(17) &  &  &                 \\ \vspace{-0.2cm}
 & & 1.7(48) &  & 1.75(35) &  &                 \\ 
Ca {\small{II}} K & 1.53(22)  &  & 1.71(17) &   &  &  \\ 
\enddata
\tablecomments{
All parameters are obtained using {\tt VoigtFit} \citep{2018arXiv180301187K} and a custom {\tt IDL} program.
$^{a}${EWs are in restframe. Errors are multiplied by 10$^2$.}
$^{b}${Total EW calculated from the fit.}
}
\end{deluxetable*}

\subsection{The Sodium and Calcium Absorption Line Components} 
\label{sec:EW}

\begin{figure}[htb!]
\epsscale{1.23}
\plotone{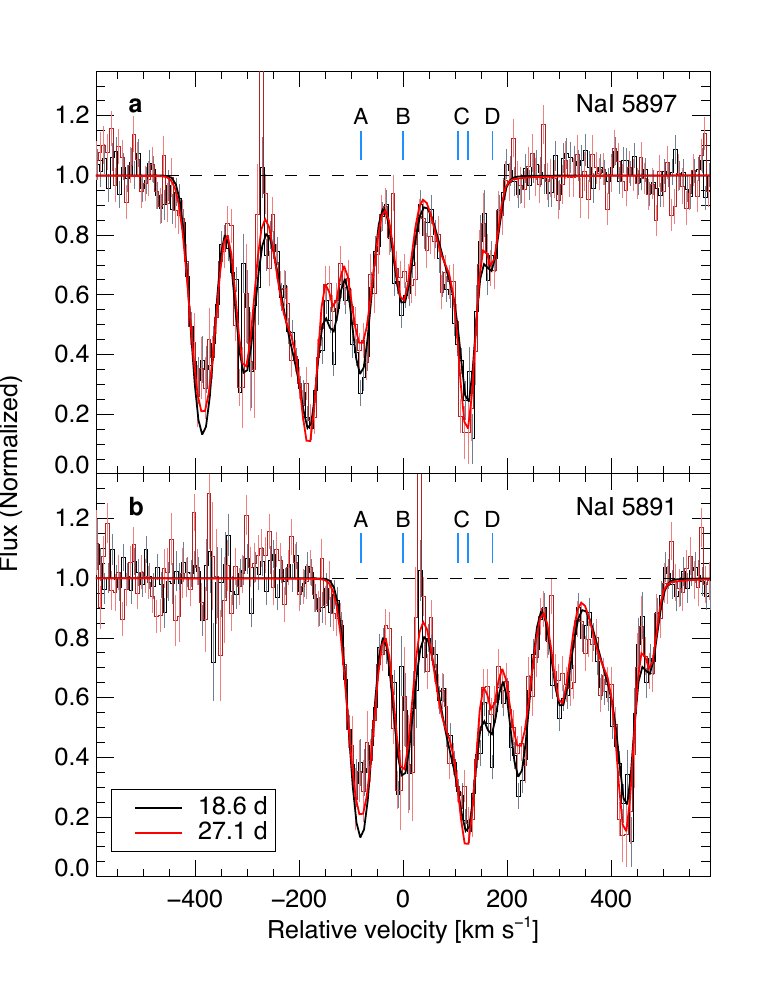}
\caption{The sodium absorption line complex of \sn. 
The panels show {\bf a,} Na~{\small{I}}~D$_1$ 
$\lambda$5897, {\bf b,} Na~{\small{I}}~D$_2$ $\lambda$5891
\AA. The data are shown as thin black and red lines 
including errors on the data for restframe epochs at 
$+$18.6 days and $+$27.1 days, respectively. The black and
red solid lines are Voigt-profile fits (Sect.~\ref{sec:NaCacomplex}) at the 
respective epochs. Five velocity components are identified
and labelled as A, B, C and D of which component C 
consists of two sub-components, C1 and C2.
The horizontal dashed line indicates 1.0 in 
normalized flux units. All velocities are given relative 
to the redshift of component B, $z$ = 0.408788. 
\label{fig:fig1}}
\end{figure}

Figure~\ref{fig:fig1} displays the remarkably complex absorption line system of 
the sodium doublet, Na~{\small{I}}~D, at both epochs. We identify 
five distinct components (A, B, C1, C2, D) for each of the 
Na~{\small{I}}~D$_1$ $\lambda$5897 \AA\ and Na~{\small{I}}~D$_2$ 
$\lambda$5891 \AA\ transitions, spanning a velocity range of about 
250 km~s$^{-1}$. The Na~{\small{I}}~D$_1$ and D$_2$ are not fully 
separated but overlap around component A of Na~{\small{I}}~D$_1$
and component D of Na~{\small{I}}~D$_2$. Separating 
Na~{\small{I}}~D$_1$ and D$_2$ results in restframe EWs of 2.29 $\pm$ 0.28 \AA\ and 3.40 $\pm$ 0.29 \AA, 
respectively. The total restframe EW of the entire Na~{\small{I}}~D 
complex is 5.2 $\pm$ 0.2 \AA. 
In an independent analysis,
\citet{2021MNRAS.502..510J} found that the sodium complex has three components and reported a total sodium restframe EW of 3.9 \AA\ and 3.3 \AA\ at the two epochs, respectively.

\begin{figure*}[htb!]
\epsscale{1.13}
\plotone{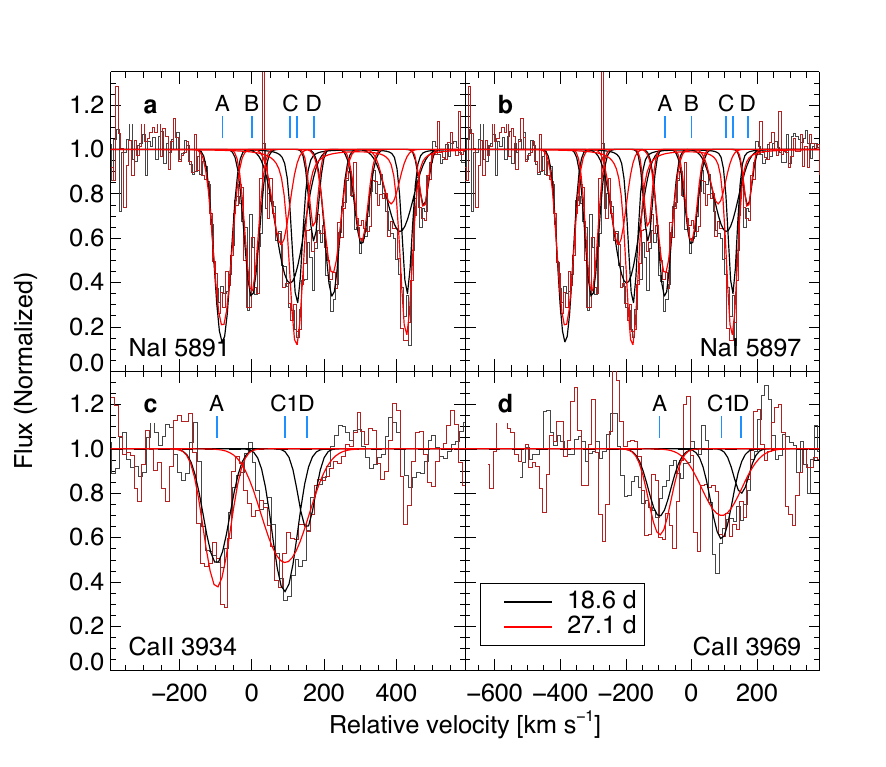}
\caption{The sodium and calcium absorption line complex of \sn. 
The panels show {\bf a,} Na~{\small{I}}~D$_2$ $\lambda$5891, 
{\bf b,} Na~{\small{I}}~D$_1$ $\lambda$5897, {\bf c,} 
Ca~{\small{II}}~K $\lambda$ 3934 and {\bf d,} 
Ca~{\small{II}}~H $\lambda$3969 \AA. 
The black and red solid lines are the individual 
Voigt-profile (Sect.~\ref{sec:NaCacomplex}) components 
from fits to the data at restframe epochs $+$18.6 days 
and $+$27.1 days, respectively.
There are five components labelled as A, B, C and D of which 
component C consists of two sub-components C1 and C2 for 
each of the Na~{\small{I}}~D$_1$ \& D$_2$ transitions and 
three corresponding 
Ca~{\small{II}}~H \& K components are labelled A, C1 and D. 
The horizontal dashed line indicates 1 in normalized flux 
units. All velocities are given relative to the redshift of 
component B, $z$ = 0.408788. 
\label{fig:fig2}}
\end{figure*}

Figure~\ref{fig:fig2} shows a comparison between the 
Na~{\small{I}}~D and Ca~{\small{II}}~H \& K absorption 
complexes. For the Ca~{\small{II}}~H \& K lines we identify 
three velocity components at similar positions as the 
Na~{\small{I}}~D components A, C1 and D. We do not identify a  
Ca~{\small{II}} counterpart to the Na~{\small{I}}~D component B. 
The Ca~{\small{II}} component C1 is broader than the Na~{\small{I}}~D C1. 
This is likely due to it being blended with 
C2. Due to the lower signal-to-noise ratio  
at the position of Ca~{\small{II}} than around 
Na~{\small{I}}~D we cannot unambiguously identify a Ca~{\small{II}} 
component C2. For the second epoch ($+$27.1 days) the Ca~{\small{II}}
components C and D are blended. The position of the component C+D remains similar to that of C1.

Through the Voigt profile fits 
we obtained $b$-values $\gtrapprox$ 20 km s$^{-1}$ for components A, 
B and C1 (Table~\ref{tab:LINDAT1}). Although the limiting instrumental 
resolution for the Na~{\small{I}}~D complex is $\sim$ 18 km s$^{-1}$,
this exceeds the maximum theoretically expected $b$-value by thermal 
broadening ($\approx$ 6 km s$^{-1}$) of the Na~{\small{I}}~D and 
Ca~{\small{II}} lines. It is also larger than the observed median 
$b$-value ($\approx$ 0.7 km s$^{-1}$) of Galactic clouds with a 
temperature of $\sim$ 80 K \citep{1994ApJ...436..152W} and the 
highest $b$-values obtained from medium or high resolution spectra 
of other SNe~Ia 
\citep{1989A&A...215...21D, 2015ApJ...799..197R, 2016A&A...592A..40F}. 
This strongly suggests that
for \sn, components A, B and C1 must 
consist of several systems that are 
unresolved in our data.
Indeed, other SNe~Ia, e.g., SN~1986G, SN~2009le and SN~2014J, have 
about 9, 8, and 18 resolved components \citep{1989A&A...215...21D, 
2013ApJ...779...38P, 2015ApJ...799..197R}, respectively. 

\begin{figure}[htb!]
\epsscale{1.175}
\plotone{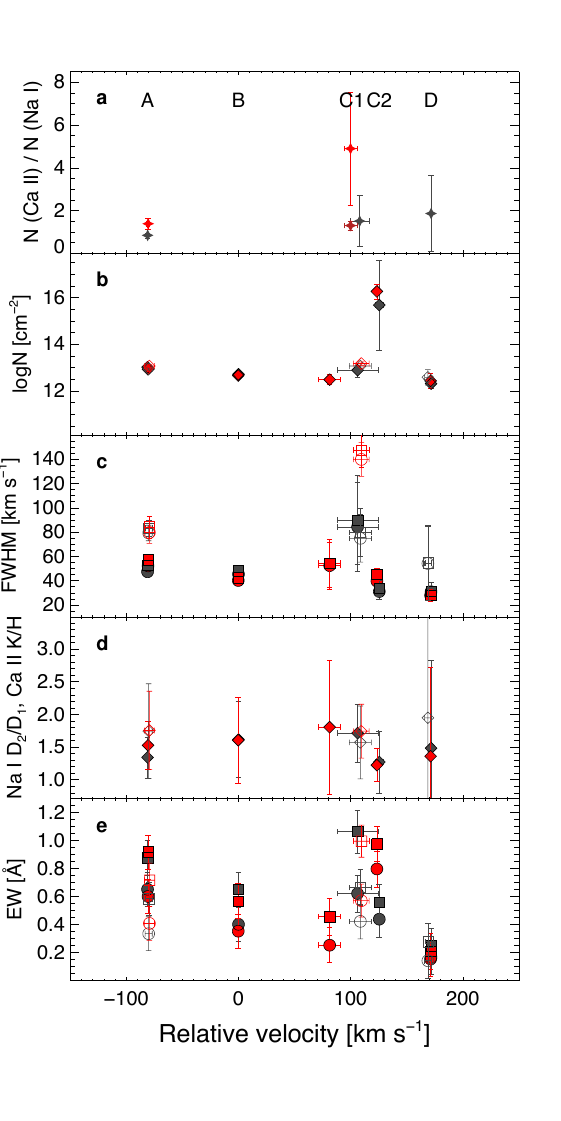}
\vspace{-1cm}
\caption{Parameters of the different Na~{\small{I}}~D and 
Ca~{\small{II}}~H \& K components. Grey and red symbols are
for epochs $+$18.6 days and $+$27.1 days, respectively. 
Filled and open symbols are for Na~{\small{I D}} and 
Ca~{\small{II}}, respectively. Square symbols mark the 
Na~{\small{I}}~D$_2$ and Ca~{\small{II}}~K transitions while 
circles mark the Na~{\small{I}}~D$_1$ and Ca~{\small{II}}~H 
transitions. All measurements are summarized in 
Tables~\ref{tab:LINDAT2} and \ref{tab:LINDAT1}. 
{\bf a,} the column density 
$N$(Ca~{\small{II}})/$N$(Na~{\small{I}}~D) ratio of component
A, C1 and D. The dark red data point ($+$27.1 days) is the 
ratio $N$(Ca~{\small{II}}(C1))/$N$(Na~{\small{I}}~D(C1+D)) at
the velocity of component C1. 
{\bf b,} the column densities. 
{\bf c,} the full width at half maximum (FWHM).
{\bf d,} the equivalent width (EW) Na~{\small{I}}~D$_2$/D$_1$
and Ca~{\small{II}}~K/H  ratios.  
{\bf e,} the EW values.  
\label{fig:fig3}}
\end{figure}

\subsection{Lack of Temporal Evolution of the Na~{\small{I}}~D EW} 
\label{sec:EW}

Figure~\ref{fig:fig3} and Table~\ref{tab:LINDAT1} show that there is 
no time variability of the measured quantities (e.g., EW, FWHM, 
column density) for any of the absorption components of 
either Na~{\small{I}}~D nor Ca~{\small{II}}~H \& K. 
On the other hand, for the same VLT-X-shooter spectra, a decline in the total Na~{\small{I}}~D EW from 3.9 \AA\ at $+$18.6 days to 3.3 \AA\ at $+$27.1 days was reported. 
These measurements are discrepant from the ones reported here. This may be due to differences in the lens and host galaxy subtraction, or possibly lack of thereof in \citet{2021MNRAS.502..510J}. Using our VLT-X-shooter spectra with neither lens nor host galaxy subtraction, we find comparable Na~{\small{I}}~D total EWs as \citet{2021MNRAS.502..510J} of 3.52 $\pm$ 0.18 \AA\ and 3.13 $\pm$ 0.2 \AA\ for the epochs $+$18.6 days and $+$27.1 days, respectively. Hence, the EWs as measured by \citet{2021MNRAS.502..510J} are a lower limit. The time variability reported by \citet{2021MNRAS.502..510J} is possibly an effect of differences in the relative contribution of the lens and host galaxy flux in the two X-shooter spectra. These were observed at different position angles with the spectroscopic slit not covering the entire source.
Indeed, time-variability of Na~{\small{I}}~D EWs is not expected during the 8.5-day (restframe) period at the epochs of our X-shooter spectra \citep{2009ApJ...693..207B, 2019ApJ...882..120W}. 

Our spectra have been corrected for lens galaxy contamination.
To test for the possible effects of Na~{\small{I}}~D absorption 
contamination from the host galaxy we model the host galaxy using the elliptical galaxy template \citep{2018MNRAS.473.4257C} that was also used to correct for lens galaxy contamination
\citep{2021MNRAS.502..510J} (see Appendix A for a discussion of the
spectral features of the lens and host galaxies). 
We subtract a scaled, linear combination of the lens and host 
spectral template at their respective redshifts and estimate the scaling factor from 
Figure~5 of \citet{2021MNRAS.502..510J}. This shows that, 
at the position of the sodium complex, the host galaxy contributes 
about one third of the total host $+$ lens spectral energy 
flux. We then repeat the absorption line analysis described above on
the lens+host subtracted spectrum and obtain a total 
Na~{\small{I}}~D EW of 5.04 $\pm$ 0.25 \AA\ ($+$18.6 days) 
and 4.8 $\pm$ 0.25 \AA\ ($+$27.1 days), which is within 
the quoted uncertainties of our measurements (Tables~\ref{tab:LINDAT1} 
and \ref{tab:LINDAT2}). 
We note that the galaxy template has prominent Na~{\small{I}}~D absorption which may 
overestimate the actual Na~{\small{I}}~D absorption from 
the host galaxy. Therefore, the effect can be considered as 
an upper limit and possible host galaxy contamination does not impact the results and 
conclusion of the paper.

\section{The Origin of the Sodium Absorption Components} 
\label{s:discussion}
%
Large EWs, similar to those measured for \sn, 
have only been measured for four other SNe~Ia in the nearby 
universe, SN~1986G \citep{1989A&A...215...21D}, SN~2003cg 
\citep{2009ApJ...693..207B}, SN~2006et \citep{2019ApJ...882..120W} and 
SN~2014J \citep{2015ApJ...799..197R}. Hence, this poses the question: 
Why does this rare, lensed SN~Ia belong to this select group of SNe~Ia?  
To address this question we assess the origin of the resolved  Na~{\small{I}}~D and Ca~{\small{II}}~H \& K components, A, B, C1 and D.

\subsection{Column Density Ratios and an Inhomegeneous Medium}

Figure~\ref{fig:fig3} and Table~\ref{tab:LINDAT1} 
show the EW ratio of individual components of the 
Na~{\small{I}} and Ca~{\small{II}}. The Na~{\small{I}}~D$_2$ 
to D$_1$ EW ratios are EW(D$_2$)/EW(D$_1$) = 1.48 $\pm$ 0.22 
(18.6 days) and 1.44 $\pm$ 0.23 (27.1 days), i.e., no variability. Values for 
the single components (A--D) are between 1.2--1.8. The 
Ca~{\small{II}} EW(K)/EW(H) ratio is $\sim$ 1.7 at both 
epochs. For homogeneous medium, EW ratios of 2 are expected 
for Na~{\small{I}}~D and Ca~{\small{II}} from their respective 
oscillator strengths ($f$-values). Ratios closer to one imply 
an inhomogeneous medium: different sub-components may only 
incompletely cover the continuum radiation source. Some such 
sub-components may be saturated, despite the fact that the 
observed absorption lines do not reach zero intensity (see 
Figures~\ref{fig:fig1} and \ref{fig:fig2}). A similar effect 
has been observed through quasar and galaxy sight lines across 
the Milky Way \citep{2012MNRAS.426.1465P} and in other SNe~Ia 
\citep[][]{1989A&A...215...21D, 2015ApJ...799..197R, 2010AJ....140.2036S}.
Indeed, the Doppler broadening parameter, $b$ $\gtrapprox$ 20 km, for components A, B and C1 implies the existence of multiple sub-components unresolved in our data and, thus, supports the inference of an inhomogenous medium.

To infer the nature of these sub-components, we use the column density ratio, $N$(Ca~{\small{II}})/$N$(Na~{\small{I}}). We obtain a ratio of 0.8 $\pm$ 0.1 for component A while C1 and D have ratios $\gtrsim$ 1 (Figure~\ref{fig:fig3}). 
Similar ratios have been measured for e.g., SN~1986G \citep{1989A&A...215...21D} and SN~2014J \citep{2015ApJ...799..197R}. 
For component B the column density ratio, $N$(Ca~{\small{II}})/$N$(Na~{\small{I}}) as well as the EW ratio, EW(Ca~{\small{II}})/EW(Na~{\small{I}}) are $\ll$~1.
Measurements of column density ratios in interstellar 
sight-lines show that $N$(Ca~{\small{II}})/$N$(Na~{\small{I}}) 
$\gtrsim$ 1 correspond to warm, high velocity (relative to the local
standard of rest) clouds in which calcium is mostly liberated from 
dust grains. Conversely, ratios $\lesssim$ 1 are typical of cold dense
molecular clouds in which large fractions of calcium are depleted 
onto dust grains \citep{1974ApJ...192...51S, 1993ApJ...411..729V}.
Consequently, the column density ratio of component A indicates 
moderate calcium depletion onto dust grains.

\subsection{Component B} 
\label{sec:EW}
It is intriguing that there is no  Ca~{\small{II}} absorption at the position of component B. 
The strong calcium  depletion of component B implies a molecular cloud origin. 

The $N$(Ca~{\small{II}})/$N$(Na~{\small{I}}) column density ratio 
is affected by variations in the ionisation conditions of the 
absorbing material. Previous studies of SNe~Ia with complex or 
time-varying Na~{\small{I}}~D have shown that due to the lower ionization 
potential and differences in the photoionization cross-sections of 
Na~{\small{I}}~D compared to Ca~{\small{II}}, Na~{\small{I}}~D is 
ionized out to radii $\approx$ 10$^{19}$ cm while Ca~{\small{II}} may
only be ionized out to shorter distances of less than 10$^{18}$ cm 
\citep{2016A&A...592A..40F, 2010A&A...514A..78P}. 
Other ionisation calculations \citep{2009ApJ...702.1157S} 
show that at low distances to SNe~Ia first Na~{\small{I}}~D 
gets ionized and Ca~{\small{II}} follows with some delay. However, 
Ca~{\small{II}} may not be affected at all if the material is at 
radii larger than $\approx$10$^{17-18}$ cm. 
In such modelled ionization scenarios, the EW of 
Na~{\small{I}}~D$_2$ is predicted to 
be less than or equal to the EW of Ca~{\small{II}}~K 
\citep{2009ApJ...702.1157S}, but can be larger if a
significant fraction of calcium is depleted onto dust grains. 
This, together with the observed time invariability 
of the Na~{\small{I}}~D components places component 
B at radii $\gtrsim$ 10$^{17-18}$ cm away from the supernova, 
which is outside the supernova dust destruction zone \citep{2019ApJ...882..120W} and, hence, supports an interstellar origin. 
Furthermore, the redshift of the host galaxy (see Sec.~\ref{sec:NaCacomplex}) coincides with component B.  

\subsection{Components A, C and D} 
\label{sec:EW}

The Na~{\small{I}}~D absorption component A is blueshifted by $\sim 80$ km s$^{-1}$. 
In comparison,
typical expansion velocities of circumstellar material 
are about 50 km s$^{-1}$ \citep{2012Sci...337..942D, 2007Sci...317..924P}. Therefore, both higher velocity blue-shifted and red-shifted 
Na~{\small{I}}~D or Ca~{\small{II}} absorption components can 
have an origin in in- and out-flowing material in the 
interstellar medium of galaxies. 
In this scenario a decline in dust 
depletion with velocity \citep{1974ApJ...192...51S}, as observed in 
\sn\ as well as in SN~1986G \citep{1989A&A...215...21D} and SN~2014J
\citep{2015ApJ...799..197R}, is expected. 
For \sn, this implies that the low, weighted average host galaxy 
$E(B-V)$ = 0.22 $\pm$ 0.04 mag \citep{2020MNRAS.491.2639D}
is in fact dominated 
by the dusty interstellar absorbers (sub-components) responsible for 
the Na~{\small{I}}~D components, i.e., A and B, while the higher velocity 
Na~{\small{I}}~D components (i.e., C and D) contribute only 
marginally to the extinction. This scenario naturally explains why the
Na~{\small{I}}~D EW of 5.2 \AA\ measured for \sn\ is much larger than expected
from existing empirical relations between dust extinction and 
Na~{\small{I}}~D EW in the Milky Way \citep{1997A&A...318..269M, 2012MNRAS.426.1465P}.

\section{Conclusion} 
\label{sec:conclusion}

In conclusion, the results demonstrate that not 
all Na~{\small{I}}~D absorbing material must contain 
dust. Hence, $E(B-V)$ measured along the line of sight 
to \sn\ is not related to the total, large 
Na~{\small{I}}~D EW of 5.2 \AA. The observed 
characteristics (column density ratios, multiple 
components and high EWs) of the Na~{\small{I}}~D and 
Ca~{\small{II}} absorption systems of \sn\ 
(Figure~\ref{fig:fig2} and \ref{fig:fig3}) reflect a 
complex, interstellar and circumgalactic in- and 
outflowing material. While the high velocity 
blue-shifted Na~{\small{I}}~D and Ca~{\small{II}} 
component A could be due to outflowing CSM, there 
is no further evidence supporting such a scenario. 
However, there is hydrogen emission of the host 
galaxy of \sn\ \citep[][see Appendix Figure~\ref{fig:fig5}]{2021MNRAS.502..510J} with 
a velocity dispersion that coincides with the velocity 
range of the Na~{\small{I}}~D. Consequently, 
such strong Na~{\small{I}}~D and Ca~{\small{II}} 
absorption systems are likely linked to gas rich star 
forming regions, young stellar populations or late 
type host galaxies \citep[][]{2012MNRAS.421.3488H}. 
Since there is evidence suggesting that the host galaxy is a massive galaxy dominated by an old stellar population, we conclude that the complex 
Na~{\small{I}}~D and Ca~{\small{II}} originates from
a star forming region in a massive galaxy.
\begin{acknowledgments}
This investigation is based on observations made with ESO Telescopes at the La Silla Paranal Observatory under programme 
ID 098.A-0648(A). This work is supported by a 
VILLUM FONDEN Young Investigator Grant (project number 25501).
This work was supported by research grants (VIL16599, VIL54489) from VILLUM FONDEN.
We thank Radek Wojtak and the anonymous referee for helpful comments and discussions. 
\end{acknowledgments}
%
%
\vspace{5mm}
\facilities{ESO-VLT(XHOOTER)}
%

\software{astropy \citep{2013A&A...558A..33A,2018AJ....156..123A},  
ESO / {\tt ESOReflex} workflow versions 2.9.2 \citep{2010SPIE.7737E..56M, 2013A&A...559A..96F}, molecfit \citep{2015A&A...576A..77S, 2015A&A...576A..78K}, custom {\tt python} programs \footnote{For program details see https://github.com/jselsing/xsh-postproc} \citep{2019A&A...623A..92S}, {\tt VoigtFit}
\citep{2018arXiv180301187K}, custom IDL programs \citep{cgdataset2024}.
%
}
%


\bibliography{oneAupdate}{}

\begin{thebibliography}{}
\expandafter\ifx\csname natexlab\endcsname\relax\def\natexlab#1{#1}\fi
\providecommand{\url}[1]{\href{#1}{#1}}
\providecommand{\dodoi}[1]{doi:~\href{http://doi.org/#1}{\nolinkurl{#1}}}
\providecommand{\doeprint}[1]{\href{http://ascl.net/#1}{\nolinkurl{http://ascl.net/#1}}}
\providecommand{\doarXiv}[1]{\href{https://arxiv.org/abs/#1}{\nolinkurl{https://arxiv.org/abs/#1}}}

\bibitem[{{Astropy Collaboration} {et~al.}(2013){Astropy Collaboration},
  {Robitaille}, {Tollerud}, {Greenfield}, {Droettboom}, {Bray}, {Aldcroft},
  {Davis}, {Ginsburg}, {Price-Whelan}, {Kerzendorf}, {Conley}, {Crighton},
  {Barbary}, {Muna}, {Ferguson}, {Grollier}, {Parikh}, {Nair}, {Unther},
  {Deil}, {Woillez}, {Conseil}, {Kramer}, {Turner}, {Singer}, {Fox}, {Weaver},
  {Zabalza}, {Edwards}, {Azalee Bostroem}, {Burke}, {Casey}, {Crawford},
  {Dencheva}, {Ely}, {Jenness}, {Labrie}, {Lim}, {Pierfederici}, {Pontzen},
  {Ptak}, {Refsdal}, {Servillat}, \& {Streicher}}]{2013A&A...558A..33A}
{Astropy Collaboration}, {Robitaille}, T.~P., {Tollerud}, E.~J., {et~al.} 2013,
  \aap, 558, A33, \dodoi{10.1051/0004-6361/201322068}

\bibitem[{{Astropy Collaboration} {et~al.}(2018){Astropy Collaboration},
  {Price-Whelan}, {Sip{\H{o}}cz}, {G{\"u}nther}, {Lim}, {Crawford}, {Conseil},
  {Shupe}, {Craig}, {Dencheva}, {Ginsburg}, {VanderPlas}, {Bradley},
  {P{\'e}rez-Su{\'a}rez}, {de Val-Borro}, {Aldcroft}, {Cruz}, {Robitaille},
  {Tollerud}, {Ardelean}, {Babej}, {Bach}, {Bachetti}, {Bakanov}, {Bamford},
  {Barentsen}, {Barmby}, {Baumbach}, {Berry}, {Biscani}, {Boquien}, {Bostroem},
  {Bouma}, {Brammer}, {Bray}, {Breytenbach}, {Buddelmeijer}, {Burke},
  {Calderone}, {Cano Rodr{\'\i}guez}, {Cara}, {Cardoso}, {Cheedella}, {Copin},
  {Corrales}, {Crichton}, {D'Avella}, {Deil}, {Depagne}, {Dietrich}, {Donath},
  {Droettboom}, {Earl}, {Erben}, {Fabbro}, {Ferreira}, {Finethy}, {Fox},
  {Garrison}, {Gibbons}, {Goldstein}, {Gommers}, {Greco}, {Greenfield},
  {Groener}, {Grollier}, {Hagen}, {Hirst}, {Homeier}, {Horton}, {Hosseinzadeh},
  {Hu}, {Hunkeler}, {Ivezi{\'c}}, {Jain}, {Jenness}, {Kanarek}, {Kendrew},
  {Kern}, {Kerzendorf}, {Khvalko}, {King}, {Kirkby}, {Kulkarni}, {Kumar},
  {Lee}, {Lenz}, {Littlefair}, {Ma}, {Macleod}, {Mastropietro}, {McCully},
  {Montagnac}, {Morris}, {Mueller}, {Mumford}, {Muna}, {Murphy}, {Nelson},
  {Nguyen}, {Ninan}, {N{\"o}the}, {Ogaz}, {Oh}, {Parejko}, {Parley}, {Pascual},
  {Patil}, {Patil}, {Plunkett}, {Prochaska}, {Rastogi}, {Reddy Janga},
  {Sabater}, {Sakurikar}, {Seifert}, {Sherbert}, {Sherwood-Taylor}, {Shih},
  {Sick}, {Silbiger}, {Singanamalla}, {Singer}, {Sladen}, {Sooley},
  {Sornarajah}, {Streicher}, {Teuben}, {Thomas}, {Tremblay}, {Turner},
  {Terr{\'o}n}, {van Kerkwijk}, {de la Vega}, {Watkins}, {Weaver}, {Whitmore},
  {Woillez}, {Zabalza}, \& {Astropy Contributors}}]{2018AJ....156..123A}
{Astropy Collaboration}, {Price-Whelan}, A.~M., {Sip{\H{o}}cz}, B.~M., {et~al.}
  2018, \aj, 156, 123, \dodoi{10.3847/1538-3881/aabc4f}

\bibitem[{{Benetti} {et~al.}(2005){Benetti}, {Cappellaro}, {Mazzali},
  {Turatto}, {Altavilla}, {Bufano}, {Elias-Rosa}, {Kotak}, {Pignata}, {Salvo},
  \& {Stanishev}}]{2005ApJ...623.1011B}
{Benetti}, S., {Cappellaro}, E., {Mazzali}, P.~A., {et~al.} 2005, \apj, 623,
  1011, \dodoi{10.1086/428608}

\bibitem[{{Blondin} {et~al.}(2009){Blondin}, {Prieto}, {Patat}, {Challis},
  {Hicken}, {Kirshner}, {Matheson}, \& {Modjaz}}]{2009ApJ...693..207B}
{Blondin}, S., {Prieto}, J.~L., {Patat}, F., {et~al.} 2009, \apj, 693, 207,
  \dodoi{10.1088/0004-637X/693/1/207}

\bibitem[{{Branch} {et~al.}(2006){Branch}, {Dang}, {Hall}, {Ketchum},
  {Melakayil}, {Parrent}, {Troxel}, {Casebeer}, {Jeffery}, \&
  {Baron}}]{2006PASP..118..560B}
{Branch}, D., {Dang}, L.~C., {Hall}, N., {et~al.} 2006, \pasp, 118, 560,
  \dodoi{10.1086/502778}

\bibitem[{{Buckley-Geer} {et~al.}(2020){Buckley-Geer}, {Lin}, {Rusu}, {Poh},
  {Palmese}, {Agnello}, {Christensen}, {Frieman}, {Shajib}, {Treu}, {Collett},
  {Birrer}, {Anguita}, {Fassnacht}, {Meylan}, {Mukherjee}, {Wong}, {Aguena},
  {Allam}, {Avila}, {Bertin}, {Bhargava}, {Brooks}, {Carnero Rosell}, {Carrasco
  Kind}, {Carretero}, {Castander}, {Costanzi}, {da Costa}, {De Vicente},
  {Desai}, {Diehl}, {Doel}, {Eifler}, {Everett}, {Flaugher}, {Fosalba},
  {Garc{\'\i}a-Bellido}, {Gaztanaga}, {Gruen}, {Gruendl}, {Gschwend},
  {Gutierrez}, {Hinton}, {Honscheid}, {James}, {Kuehn}, {Kuropatkin}, {Maia},
  {Marshall}, {Melchior}, {Menanteau}, {Miquel}, {Ogand o}, {Paz-Chinch{\'o}n},
  {Plazas}, {Sanchez}, {Scarpine}, {Schubnell}, {Serrano}, {Sevilla-Noarbe},
  {Smith}, {Soares-Santos}, {Suchyta}, {Swanson}, {Tarle}, {Tucker}, {Varga},
  \& {DES Collaboration}}]{2020MNRAS.498.3241B}
{Buckley-Geer}, E.~J., {Lin}, H., {Rusu}, C.~E., {et~al.} 2020, \mnras, 498,
  3241, \dodoi{10.1093/mnras/staa2563}

\bibitem[{{Cano} {et~al.}(2018){Cano}, {Selsing}, {Hjorth}, {de Ugarte
  Postigo}, {Christensen}, {Gall}, \& {Kann}}]{2018MNRAS.473.4257C}
{Cano}, Z., {Selsing}, J., {Hjorth}, J., {et~al.} 2018, \mnras, 473, 4257,
  \dodoi{10.1093/mnras/stx2624}

\bibitem[{{Clark} {et~al.}(2021){Clark}, {Maguire}, {Bulla}, {Galbany},
  {Sullivan}, {Anderson}, \& {Smartt}}]{2021MNRAS.507.4367C}
{Clark}, P., {Maguire}, K., {Bulla}, M., {et~al.} 2021, \mnras, 507, 4367,
  \dodoi{10.1093/mnras/stab2038}

\bibitem[{{Dhawan} {et~al.}(2020){Dhawan}, {Johansson}, {Goobar}, {Amanullah},
  {M{\"o}rtsell}, {Cenko}, {Cooray}, {Fox}, {Goldstein}, {Kalender},
  {Kasliwal}, {Kulkarni}, {Lee}, {Nayyeri}, {Nugent}, {Ofek}, \&
  {Quimby}}]{2020MNRAS.491.2639D}
{Dhawan}, S., {Johansson}, J., {Goobar}, A., {et~al.} 2020, \mnras, 491, 2639,
  \dodoi{10.1093/mnras/stz2965}

\bibitem[{{Dilday} {et~al.}(2012){Dilday}, {Howell}, {Cenko}, {Silverman},
  {Nugent}, {Sullivan}, {Ben-Ami}, {Bildsten}, {Bolte}, {Endl}, {Filippenko},
  {Gnat}, {Horesh}, {Hsiao}, {Kasliwal}, {Kirkman}, {Maguire}, {Marcy},
  {Moore}, {Pan}, {Parrent}, {Podsiadlowski}, {Quimby}, {Sternberg}, {Suzuki},
  {Tytler}, {Xu}, {Bloom}, {Gal-Yam}, {Hook}, {Kulkarni}, {Law}, {Ofek},
  {Polishook}, \& {Poznanski}}]{2012Sci...337..942D}
{Dilday}, B., {Howell}, D.~A., {Cenko}, S.~B., {et~al.} 2012, Science, 337,
  942, \dodoi{10.1126/science.1219164}

\bibitem[{{D'Odorico} {et~al.}(1989){D'Odorico}, {di Serego Alighieri},
  {Pettini}, {Magain}, {Nissen}, \& {Panagia}}]{1989A&A...215...21D}
{D'Odorico}, S., {di Serego Alighieri}, S., {Pettini}, M., {et~al.} 1989, \aap,
  215, 21

\bibitem[{{D'Odorico} {et~al.}(2006){D'Odorico}, {Dekker}, {Mazzoleni},
  {Vernet}, {Guinouard}, {Groot}, {Hammer}, {Rasmussen}, {Kaper}, {Navarro},
  {Pallavicini}, {Peroux}, \& {Zerbi}}]{2006SPIE.6269E..98D}
{D'Odorico}, S., {Dekker}, H., {Mazzoleni}, R., {et~al.} 2006, in Society of
  Photo-Optical Instrumentation Engineers (SPIE) Conference Series, Vol. 6269,
  Society of Photo-Optical Instrumentation Engineers (SPIE) Conference Series,
  \dodoi{10.1117/12.672969}

\bibitem[{{Ferretti} {et~al.}(2016){Ferretti}, {Amanullah}, {Goobar},
  {Johansson}, {Vreeswijk}, {Butler}, {Cao}, {Cenko}, {Doran}, {Filippenko},
  {Freeland }, {Hosseinzadeh}, {Howell}, {Lundqvist}, {Mattila}, {Nordin},
  {Nugent}, {Petrushevska}, {Valenti}, {Vogt}, \&
  {Wozniak}}]{2016A&A...592A..40F}
{Ferretti}, R., {Amanullah}, R., {Goobar}, A., {et~al.} 2016, \aap, 592, A40,
  \dodoi{10.1051/0004-6361/201628351}

\bibitem[{{Foley} {et~al.}(2012){Foley}, {Simon}, {Burns}, {Gal-Yam}, {Hamuy},
  {Kirshner}, {Morrell}, {Phillips}, {Shields}, \&
  {Sternberg}}]{2012ApJ...752..101F}
{Foley}, R.~J., {Simon}, J.~D., {Burns}, C.~R., {et~al.} 2012, \apj, 752, 101,
  \dodoi{10.1088/0004-637X/752/2/101}

\bibitem[{{Freudling} {et~al.}(2013){Freudling}, {Romaniello}, {Bramich},
  {Ballester}, {Forchi}, {Garc{\'\i}a-Dabl{\'o}}, {Moehler}, \&
  {Neeser}}]{2013A&A...559A..96F}
{Freudling}, W., {Romaniello}, M., {Bramich}, D.~M., {et~al.} 2013, \aap, 559,
  A96, \dodoi{10.1051/0004-6361/201322494}

\bibitem[{{Gall} {et~al.}(2024){Gall}, {Hjorth}, {Christensen}, {Izzo}, \&
  {Selsing}}]{cgdataset2024}
{Gall}, C., {Hjorth}, J., {Christensen}, L., {Izzo}, L., \& {Selsing}, J. 2024,
  {Sodium towards SN~2016geu},  University of Copenhagen,
  \dodoi{10.17894/UCPH.940A6E89-B8FF-4265-AF45-BFBE6AD06D0C}

\bibitem[{{Goobar} {et~al.}(2017){Goobar}, {Amanullah}, {Kulkarni}, {Nugent},
  {Johansson}, {Steidel}, {Law}, {M{\"o}rtsell}, {Quimby}, {Blagorodnova},
  {Brand eker}, {Cao}, {Cooray}, {Ferretti}, {Fremling}, {Hangard}, {Kasliwal},
  {Kupfer}, {Lunnan}, {Masci}, {Miller}, {Nayyeri}, {Neill}, {Ofek},
  {Papadogiannakis}, {Petrushevska}, {Ravi}, {Sollerman}, {Sullivan}, {Taddia},
  {Walters}, {Wilson}, {Yan}, \& {Yaron}}]{2017Sci...356..291G}
{Goobar}, A., {Amanullah}, R., {Kulkarni}, S.~R., {et~al.} 2017, Science, 356,
  291, \dodoi{10.1126/science.aal2729}

\bibitem[{{Graham} {et~al.}(2015){Graham}, {Valenti}, {Fulton}, {Weiss},
  {Shen}, {Kelly}, {Zheng}, {Filippenko}, {Marcy}, {Howell}, {Burt}, \&
  {Rivera}}]{2015ApJ...801..136G}
{Graham}, M.~L., {Valenti}, S., {Fulton}, B.~J., {et~al.} 2015, \apj, 801, 136,
  \dodoi{10.1088/0004-637X/801/2/136}

\bibitem[{{Hillebrandt} \& {Niemeyer}(2000)}]{2000ARA&A..38..191H}
{Hillebrandt}, W., \& {Niemeyer}, J.~C. 2000, \araa, 38, 191,
  \dodoi{10.1146/annurev.astro.38.1.191}

\bibitem[{{H{\"o}flich} {et~al.}(2011){H{\"o}flich}, {Kumar}, \&
  {Wheeler}}]{2011cetd.book.....H}
{H{\"o}flich}, P., {Kumar}, P., \& {Wheeler}, J.~C. 2011, {Cosmic Explosions in
  Three Dimensions}

\bibitem[{{Hopkins} {et~al.}(2012){Hopkins}, {Quataert}, \&
  {Murray}}]{2012MNRAS.421.3488H}
{Hopkins}, P.~F., {Quataert}, E., \& {Murray}, N. 2012, \mnras, 421, 3488,
  \dodoi{10.1111/j.1365-2966.2012.20578.x}

\bibitem[{{Horne}(1986)}]{1986PASP...98..609H}
{Horne}, K. 1986, \pasp, 98, 609, \dodoi{10.1086/131801}

\bibitem[{{Hoyle} \& {Fowler}(1960)}]{1960ApJ...132..565H}
{Hoyle}, F., \& {Fowler}, W.~A. 1960, \apj, 132, 565, \dodoi{10.1086/146963}

\bibitem[{{Iben} \& {Tutukov}(1984)}]{1984ApJS...54..335I}
{Iben}, I., J., \& {Tutukov}, A.~V. 1984, \apjs, 54, 335,
  \dodoi{10.1086/190932}

\bibitem[{{Johansson} {et~al.}(2021){Johansson}, {Goobar}, {Price}, {Sagu{\'e}s
  Carracedo}, {Della Bruna}, {Nugent}, {Dhawan}, {M{\"o}rtsell},
  {Papadogiannakis}, {Amanullah}, {Goldstein}, {Cenko}, {De}, {Dugas},
  {Kasliwal}, {Kulkarni}, \& {Lunnan}}]{2021MNRAS.502..510J}
{Johansson}, J., {Goobar}, A., {Price}, S.~H., {et~al.} 2021, \mnras, 502, 510,
  \dodoi{10.1093/mnras/staa3829}

\bibitem[{{Kausch} {et~al.}(2015){Kausch}, {Noll}, {Smette}, {Kimeswenger},
  {Barden}, {Szyszka}, {Jones}, {Sana}, {Horst}, \&
  {Kerber}}]{2015A&A...576A..78K}
{Kausch}, W., {Noll}, S., {Smette}, A., {et~al.} 2015, \aap, 576, A78,
  \dodoi{10.1051/0004-6361/201423909}

\bibitem[{{Krogager}(2018)}]{2018arXiv180301187K}
{Krogager}, J.-K. 2018, arXiv e-prints, arXiv:1803.01187.
\newblock \doarXiv{1803.01187}

\bibitem[{{Livio} \& {Mazzali}(2018)}]{2018PhR...736....1L}
{Livio}, M., \& {Mazzali}, P. 2018, \physrep, 736, 1,
  \dodoi{10.1016/j.physrep.2018.02.002}

\bibitem[{{Maguire} {et~al.}(2013){Maguire}, {Sullivan}, {Patat}, {Gal-Yam},
  {Hook}, {Dhawan}, {Howell}, {Mazzali}, {Nugent}, {Pan}, {Podsiadlowski},
  {Simon}, {Sternberg}, {Valenti}, {Baltay}, {Bersier}, {Blagorodnova}, {Chen},
  {Ellman}, {Feindt}, {F{\"o}rster}, {Fraser}, {Gonz{\'a}lez-Gait{\'a}n},
  {Graham}, {Guti{\'e}rrez}, {Hachinger}, {Hadjiyska}, {Inserra}, {Knapic},
  {Laher}, {Leloudas}, {Margheim}, {McKinnon}, {Molinaro}, {Morrell}, {Ofek},
  {Rabinowitz}, {Rest}, {Sand}, {Smareglia}, {Smartt}, {Taddia}, {Walker},
  {Walton}, \& {Young}}]{2013MNRAS.436..222M}
{Maguire}, K., {Sullivan}, M., {Patat}, F., {et~al.} 2013, \mnras, 436, 222,
  \dodoi{10.1093/mnras/stt1586}

\bibitem[{{Modigliani} {et~al.}(2010){Modigliani}, {Goldoni}, {Royer},
  {Haigron}, {Guglielmi}, {Fran{\c c}ois}, {Horrobin}, {Bristow}, {Vernet},
  {Moehler}, {Kerber}, {Ballester}, {Mason}, \&
  {Christensen}}]{2010SPIE.7737E..56M}
{Modigliani}, A., {Goldoni}, P., {Royer}, F., {et~al.} 2010, in Society of
  Photo-Optical Instrumentation Engineers (SPIE) Conference Series, Vol. 7737,
  Society of Photo-Optical Instrumentation Engineers (SPIE) Conference Series,
  \dodoi{10.1117/12.857211}

\bibitem[{{More} {et~al.}(2017){More}, {Suyu}, {Oguri}, {More}, \&
  {Lee}}]{2017ApJ...835L..25M}
{More}, A., {Suyu}, S.~H., {Oguri}, M., {More}, S., \& {Lee}, C.-H. 2017,
  \apjl, 835, L25, \dodoi{10.3847/2041-8213/835/2/L25}

\bibitem[{{Morton}(1991)}]{1991ApJS...77..119M}
{Morton}, D.~C. 1991, \apjs, 77, 119, \dodoi{10.1086/191601}

\bibitem[{{Munari} \& {Zwitter}(1997)}]{1997A&A...318..269M}
{Munari}, U., \& {Zwitter}, T. 1997, \aap, 318, 269

\bibitem[{{Nomoto}(1982)}]{1982ApJ...253..798N}
{Nomoto}, K. 1982, \apj, 253, 798, \dodoi{10.1086/159682}

\bibitem[{{Patat} {et~al.}(2011){Patat}, {Chugai}, {Podsiadlowski}, {Mason},
  {Melo}, \& {Pasquini}}]{2011A&A...530A..63P}
{Patat}, F., {Chugai}, N.~N., {Podsiadlowski}, P., {et~al.} 2011, \aap, 530,
  A63, \dodoi{10.1051/0004-6361/201116865}

\bibitem[{{Patat} {et~al.}(2010){Patat}, {Cox}, {Parrent}, \&
  {Branch}}]{2010A&A...514A..78P}
{Patat}, F., {Cox}, N.~L.~J., {Parrent}, J., \& {Branch}, D. 2010, \aap, 514,
  A78, \dodoi{10.1051/0004-6361/200913959}

\bibitem[{{Patat} {et~al.}(2007){Patat}, {Chandra}, {Chevalier}, {Justham},
  {Podsiadlowski}, {Wolf}, {Gal-Yam}, {Pasquini}, {Crawford}, {Mazzali},
  {Pauldrach}, {Nomoto}, {Benetti}, {Cappellaro}, {Elias-Rosa}, {Hillebrandt},
  {Leonard}, {Pastorello}, {Renzini}, {Sabbadin}, {Simon}, \&
  {Turatto}}]{2007Sci...317..924P}
{Patat}, F., {Chandra}, P., {Chevalier}, R., {et~al.} 2007, Science, 317, 924,
  \dodoi{10.1126/science.1143005}

\bibitem[{{Phillips} {et~al.}(2013){Phillips}, {Simon}, {Morrell}, {Burns},
  {Cox}, {Foley}, {Karakas}, {Patat}, {Sternberg}, {Williams}, {Gal-Yam},
  {Hsiao}, {Leonard}, {Persson}, {Stritzinger}, {Thompson}, {Campillay},
  {Contreras}, {Folatelli}, {Freedman}, {Hamuy}, {Roth}, {Shields}, {Suntzeff},
  {Chomiuk}, {Ivans}, {Madore}, {Penprase}, {Perley}, {Pignata}, {Preston}, \&
  {Soderberg}}]{2013ApJ...779...38P}
{Phillips}, M.~M., {Simon}, J.~D., {Morrell}, N., {et~al.} 2013, \apj, 779, 38,
  \dodoi{10.1088/0004-637X/779/1/38}

\bibitem[{{Poznanski} {et~al.}(2012){Poznanski}, {Prochaska}, \&
  {Bloom}}]{2012MNRAS.426.1465P}
{Poznanski}, D., {Prochaska}, J.~X., \& {Bloom}, J.~S. 2012, \mnras, 426, 1465,
  \dodoi{10.1111/j.1365-2966.2012.21796.x}

\bibitem[{{Raskin} \& {Kasen}(2013)}]{2013ApJ...772....1R}
{Raskin}, C., \& {Kasen}, D. 2013, \apj, 772, 1,
  \dodoi{10.1088/0004-637X/772/1/1}

\bibitem[{{Ritchey} {et~al.}(2015){Ritchey}, {Welty}, {Dahlstrom}, \&
  {York}}]{2015ApJ...799..197R}
{Ritchey}, A.~M., {Welty}, D.~E., {Dahlstrom}, J.~A., \& {York}, D.~G. 2015,
  \apj, 799, 197, \dodoi{10.1088/0004-637X/799/2/197}

\bibitem[{{Schlafly} \& {Finkbeiner}(2011)}]{2011ApJ...737..103S}
{Schlafly}, E.~F., \& {Finkbeiner}, D.~P. 2011, \apj, 737, 103,
  \dodoi{10.1088/0004-637X/737/2/103}

\bibitem[{{Selsing} {et~al.}(2019){Selsing}, {Malesani}, {Goldoni}, {Fynbo},
  {Kr{\"u}hler}, {Antonelli}, {Arabsalmani}, {Bolmer}, {Cano}, {Christensen},
  {Covino}, {D'Avanzo}, {D'Elia}, {De Cia}, {de Ugarte Postigo}, {Flores},
  {Friis}, {Gomboc}, {Greiner}, {Groot}, {Hammer}, {Hartoog}, {Heintz},
  {Hjorth}, {Jakobsson}, {Japelj}, {Kann}, {Kaper}, {Ledoux}, {Leloudas},
  {Levan}, {Maiorano}, {Melandri}, {Milvang-Jensen}, {Palazzi}, {Palmerio},
  {Perley}, {Pian}, {Piranomonte}, {Pugliese}, {S{\'a}nchez-Ram{\'\i}rez},
  {Savaglio}, {Schady}, {Schulze}, {Sollerman}, {Sparre}, {Tagliaferri},
  {Tanvir}, {Th{\"o}ne}, {Vergani}, {Vreeswijk}, {Watson}, {Wiersema},
  {Wijers}, {Xu}, \& {Zafar}}]{2019A&A...623A..92S}
{Selsing}, J., {Malesani}, D., {Goldoni}, P., {et~al.} 2019, \aap, 623, A92,
  \dodoi{10.1051/0004-6361/201832835}

\bibitem[{{Siluk} \& {Silk}(1974)}]{1974ApJ...192...51S}
{Siluk}, R.~S., \& {Silk}, J. 1974, \apj, 192, 51, \dodoi{10.1086/153033}

\bibitem[{{Simon} {et~al.}(2009){Simon}, {Gal-Yam}, {Gnat}, {Quimby},
  {Ganeshalingam}, {Silverman}, {Blondin}, {Li}, {Filippenko}, {Wheeler},
  {Kirshner}, {Patat}, {Nugent}, {Foley}, {Vogt}, {Butler}, {Peek},
  {Rosolowsky}, {Herczeg}, {Sauer}, \& {Mazzali}}]{2009ApJ...702.1157S}
{Simon}, J.~D., {Gal-Yam}, A., {Gnat}, O., {et~al.} 2009, \apj, 702, 1157,
  \dodoi{10.1088/0004-637X/702/2/1157}

\bibitem[{{Smette} {et~al.}(2015){Smette}, {Sana}, {Noll}, {Horst}, {Kausch},
  {Kimeswenger}, {Barden}, {Szyszka}, {Jones}, {Gallenne}, {Vinther},
  {Ballester}, \& {Taylor}}]{2015A&A...576A..77S}
{Smette}, A., {Sana}, H., {Noll}, S., {et~al.} 2015, \aap, 576, A77,
  \dodoi{10.1051/0004-6361/201423932}

\bibitem[{{Soker} {et~al.}(2013){Soker}, {Kashi}, {Garc{\'\i}a-Berro},
  {Torres}, \& {Camacho}}]{2013MNRAS.431.1541S}
{Soker}, N., {Kashi}, A., {Garc{\'\i}a-Berro}, E., {Torres}, S., \& {Camacho},
  J. 2013, \mnras, 431, 1541, \dodoi{10.1093/mnras/stt271}

\bibitem[{{Sternberg} {et~al.}(2011){Sternberg}, {Gal-Yam}, {Simon}, {Leonard},
  {Quimby}, {Phillips}, {Morrell}, {Thompson}, {Ivans}, {Marshall},
  {Filippenko}, {Marcy}, {Bloom}, {Patat}, {Foley}, {Yong}, {Penprase},
  {Beeler}, {Allende Prieto}, \& {Stringfellow}}]{2011Sci...333..856S}
{Sternberg}, A., {Gal-Yam}, A., {Simon}, J.~D., {et~al.} 2011, Science, 333,
  856, \dodoi{10.1126/science.1203836}

\bibitem[{{Stritzinger} {et~al.}(2010){Stritzinger}, {Burns}, {Phillips},
  {Folatelli}, {Krisciunas}, {Kattner}, {Persson}, {Boldt}, {Campillay},
  {Contreras}, {Krzeminski}, {Morrell}, {Salgado}, {Freedman}, {Hamuy},
  {Madore}, {Roth}, \& {Suntzeff}}]{2010AJ....140.2036S}
{Stritzinger}, M., {Burns}, C.~R., {Phillips}, M.~M., {et~al.} 2010, \aj, 140,
  2036, \dodoi{10.1088/0004-6256/140/6/2036}

\bibitem[{{Terlevich} {et~al.}(1990){Terlevich}, {D{\'\i}az}, \&
  {Terlevich}}]{1990RMxAA..21..218T}
{Terlevich}, E., {D{\'\i}az}, A.~I., \& {Terlevich}, R. 1990, \rmxaa, 21, 218

\bibitem[{{Vallerga} {et~al.}(1993){Vallerga}, {Vedder}, {Craig}, \&
  {Welsh}}]{1993ApJ...411..729V}
{Vallerga}, J.~V., {Vedder}, P.~W., {Craig}, N., \& {Welsh}, B.~Y. 1993, \apj,
  411, 729, \dodoi{10.1086/172875}

\bibitem[{{Vernet} {et~al.}(2011){Vernet}, {Dekker}, {D'Odorico}, {Kaper},
  {Kjaergaard}, {Hammer}, {Randich}, {Zerbi}, {Groot}, {Hjorth}, {Guinouard},
  {Navarro}, {Adolfse}, {Albers}, {Amans}, {Andersen}, {Andersen}, {Binetruy},
  {Bristow}, {Castillo}, {Chemla}, {Christensen}, {Conconi}, {Conzelmann},
  {Dam}, {de Caprio}, {de Ugarte Postigo}, {Delabre}, {di Marcantonio},
  {Downing}, {Elswijk}, {Finger}, {Fischer}, {Flores}, {Fran{\c{c}}ois},
  {Goldoni}, {Guglielmi}, {Haigron}, {Hanenburg}, {Hendriks}, {Horrobin},
  {Horville}, {Jessen}, {Kerber}, {Kern}, {Kiekebusch}, {Kleszcz}, {Klougart},
  {Kragt}, {Larsen}, {Lizon}, {Lucuix}, {Mainieri}, {Manuputy}, {Martayan},
  {Mason}, {Mazzoleni}, {Michaelsen}, {Modigliani}, {Moehler}, {M{\o}ller},
  {Norup S{\o}rensen}, {N{\o}rregaard}, {P{\'e}roux}, {Patat}, {Pena}, {Pragt},
  {Reinero}, {Rigal}, {Riva}, {Roelfsema}, {Royer}, {Sacco}, {Santin},
  {Schoenmaker}, {Spano}, {Sweers}, {Ter Horst}, {Tintori}, {Tromp}, {van
  Dael}, {van der Vliet}, {Venema}, {Vidali}, {Vinther}, {Vola}, {Winters},
  {Wistisen}, {Wulterkens}, \& {Zacchei}}]{2011A&A...536A.105V}
{Vernet}, J., {Dekker}, H., {D'Odorico}, S., {et~al.} 2011, \aap, 536, A105,
  \dodoi{10.1051/0004-6361/201117752}

\bibitem[{{Wang} {et~al.}(2019){Wang}, {Chen}, {Wang}, {Hu}, {Xi}, {Yang},
  {Zhao}, \& {Li}}]{2019ApJ...882..120W}
{Wang}, X., {Chen}, J., {Wang}, L., {et~al.} 2019, \apj, 882, 120,
  \dodoi{10.3847/1538-4357/ab26b5}

\bibitem[{{Webbink}(1984)}]{1984ApJ...277..355W}
{Webbink}, R.~F. 1984, \apj, 277, 355, \dodoi{10.1086/161701}

\bibitem[{{Welty} \& {Hobbs}(2001)}]{2001ApJS..133..345W}
{Welty}, D.~E., \& {Hobbs}, L.~M. 2001, \apjs, 133, 345, \dodoi{10.1086/320354}

\bibitem[{{Welty} {et~al.}(1994){Welty}, {Hobbs}, \&
  {Kulkarni}}]{1994ApJ...436..152W}
{Welty}, D.~E., {Hobbs}, L.~M., \& {Kulkarni}, V.~P. 1994, \apj, 436, 152,
  \dodoi{10.1086/174889}

\bibitem[{{Welty} {et~al.}(2014){Welty}, {Ritchey}, {Dahlstrom}, \&
  {York}}]{2014ApJ...792..106W}
{Welty}, D.~E., {Ritchey}, A.~M., {Dahlstrom}, J.~A., \& {York}, D.~G. 2014,
  \apj, 792, 106, \dodoi{10.1088/0004-637X/792/2/106}

\bibitem[{{Whelan} \& {Iben}(1973)}]{1973ApJ...186.1007W}
{Whelan}, J., \& {Iben}, Icko, J. 1973, \apj, 186, 1007, \dodoi{10.1086/152565}

\end{thebibliography}
\bibliographystyle{aasjournal}
\appendix
\section{Emission and absorption lines in the host and lens galaxy} 
\label{sec:origin}
In this appendix we provide an analysis of spectral features relating to the host and lens galaxies. These support our assumption of these galaxies being massive, dominated by old stellar populations.

\begin{figure}[htb!]
\epsscale{1.2}
\plotone{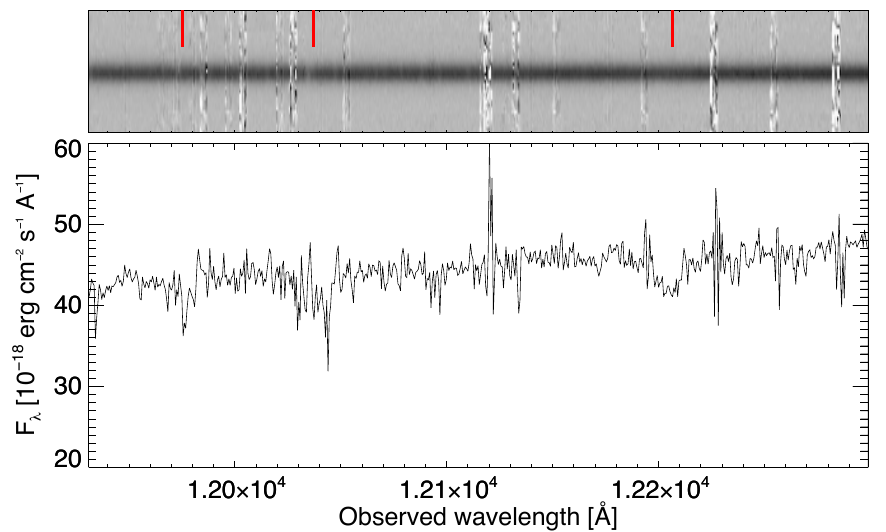}
\caption{The Ca~{\small{II}} near infrared triplet in the host 
galaxy.  
{\it Upper panel:} the co-added VLT/X-shooter 2D-spectrum of epochs 
$+$18.6 days and $+$27.1 days. The expected Ca~{\small{II}} 
triplet at $\lambda\lambda$8498, 8542 and $\lambda$8662 \AA\ 
absorption lines are indicated as red solid lines at the 
redshift of the Na~{\small{I}}~D component B. 
{\it Lower panel:} the co-added VLT/X-shooter 1D-spectrum of epochs 
$+$18.6 days and $+$27.1 days.
\label{fig:fig4}}
\end{figure}

Figure~\ref{fig:fig4} shows weak absorption of the Ca~{\small{II}} 
$\lambda $8662 \AA\ line, the strongest line of the Ca~{\small{II}} 
near infrared triplet which coincides with the redshift of the 
Na~{\small{I}}~D component B. We measure a FWHM of about 220 km s$^{-1}$. 
The Ca {\small{II}} line is observed at a wavelength of 12206.5 \AA\
right in between two OH molecular skylines, but is not affected by 
telluric absorption lines. 
The Ca~{\small{II}} NIR triplet is a prominent feature observed in  
galaxies and originates in the photospheres of stars 
\citep{1990RMxAA..21..218T}. 

\begin{figure}[htb!]
\epsscale{1.2}
\plotone{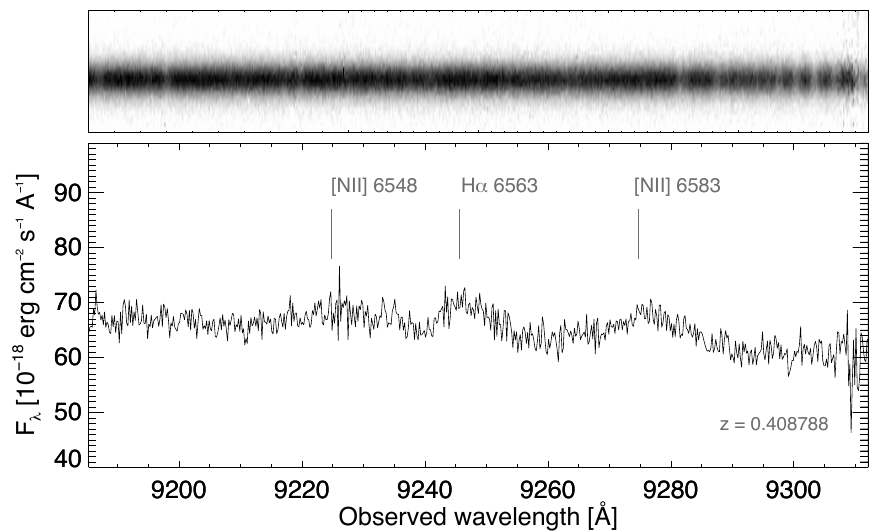}
\caption{
The H$\alpha$ emission in the host 
galaxy.  
{\it Upper panel:} the co-added VLT/X-shooter 2D-spectrum of epochs 
$+$18.6 days and $+$27.1 days. 
{\it Lower panel:} the co-added VLT/X-shooter 1D-spectrum of epochs 
$+$18.6 days and $+$27.1 days. The grey annotations mark [N~{\small{II}}] at $\lambda\lambda$ 6548.06 and 6583.45 \AA\ and  H$\alpha$ emission at the redshift of the Na~{\small{I}}~D component B.
\label{fig:fig5}}
\end{figure}

Figure~\ref{fig:fig5} shows that there is weak 
[N~{\small{II}}] $\lambda$6583 \AA\ and H$\alpha$ emission, as also noticed by
\citet{2021MNRAS.502..510J}, in the host galaxy with a FWHM $\sim$ 200
km s$^{-1}$, consistent with that of the Ca~{\small{II}} $\lambda $8662 
\AA\ line. Both the lens and host galaxy H$\alpha$/[N~{\small{II}}] flux ratios are $\sim$ 1, which is indicative of a massive galaxy and supersolar metallicity. 
However, we do not detect [O~{\small{III}}] $\lambda\lambda$ 4959, 5007 \AA, which is a typical signature for star-forming galaxies. 

\begin{figure}[htb!]
\epsscale{1.2}
\plotone{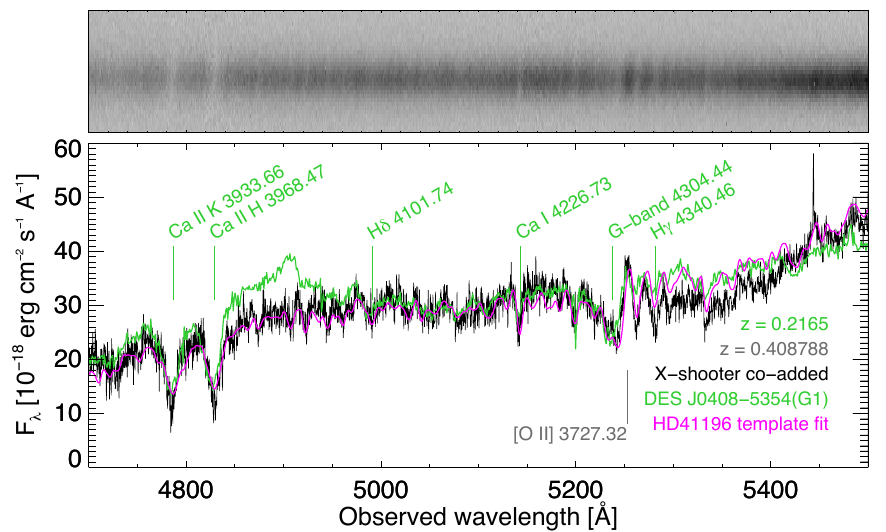}
\caption{Disentangling lens and host galaxy.
{\it Upper panel:} our co-added VLT/X-shooter  2D-spectrum of epochs 
$+$18.6 days and $+$27.1 days.
{\it Lower panel:} our co-added VLT/X-shooter 1D-spectrum of epochs 
$+$18.6 days and $+$27.1 days (black). The green spectrum is 
the massive lens galaxy DES J0408-5354 G1 
\citep{2020MNRAS.498.3241B} spectrum redshifted 
to match the redshift (z = 0.2165) of the lens galaxy of 
\sn\ and normalized in flux to match our co-added 
VLT/X-shooter spectrum. In magenta, the best fit X-shooter 
spectrum of the star HD 41196 convolved to a velocity 
dispersion of 141 $\pm$ 5 km s$^{-1}$ at the redshift of the 
lens galaxy ($z$ = 0.2164). The grey annotation illustrates 
the position of [O {\small{II}}] at the redshift 
$z$ = 0.408788 of the host galaxy of \sn\ as 
suggested by \citet{2021MNRAS.502..510J}.
\label{fig:fig6}}
\end{figure}

Figure~\ref{fig:fig6} shows a conspicuous emission 
feature at an observed wavelength of $\lambda$5250 \AA, which has 
been suggested to be [O~{\small{II}}] $\lambda$3727 \AA\  
\citep{2021MNRAS.502..510J} at a redshift, $z$ = 0.4087 of the host 
galaxy. However, it is within a complex spectral 
region. To verify its nature, we co-added our two X-shooter spectra to 
increase the signal-to-noise ratio and compared this co-added 
spectrum to a similar massive lens galaxy spectrum. The 
conspicuous emission is consistent with spectral features typical 
of a massive galaxy, i.e., molecular CH absorption at $\sim 
\lambda$4304 \AA\ ($G$-band) at a redshift of the lens galaxy $z$ = 
0.2164. Fitting our co-added X-shooter spectrum using the X-shooter 
spectrum of the star HD41196 convolved with a velocity dispersion 
of 141 $\pm$ 5 km s$^{-1}$ at the redshift of the lens galaxy can as 
well explain the spectral features in the wavelength range 
5200--5300 \AA. 
Nevertheless, weak [O~{\small{II}}] may be possible. 
Signatures of an old stellar population in the host galaxy 
include weak $G$-band or Mg~{\small{I}} triplet absorption, also typical of a massive galaxy.

In some high resolution spectra of SNe~Ia that exhibit strong Na~{\small{I}}~D there is absorption from diffuse interstellar bands (DIBs) such as the DIB at $\lambda$5780 \AA\ \citep{2013ApJ...779...38P,2015ApJ...801..136G}. However, for \sn\  
the DIB $\lambda$5780 \AA\ is not detected \citep{2021MNRAS.502..510J}. Furthermore, we do not detect other DIBs 
at $\lambda\lambda$ 5797, 6196, 6203, 6270, 6284, 6379, 6614 and 6661 \AA, molecular features such as CH$^{+}$ $\lambda\lambda$3957 and 4232 \AA\, the CN violet band $\lambda\sim$3874-3880 \AA\ or potassium, 
K~{\small{I}} $\lambda\lambda$7667 and 7701 \AA\ absorption lines. 
These elements and molecules are less abundant than sodium 
or calcium, but some have been detected in sight lines towards SN 2014J and 1986G \citep{1989A&A...215...21D, 2014ApJ...792..106W, 2015ApJ...799..197R}. Assuming that the relation between 
interstellar Na and K column densities of 
$N$(Na~{\small{II}})/$N$(K~{\small{I}}) $\approx$ 85 holds 
\citep{2001ApJS..133..345W}, it is expected that we do not detect
the K~{\small{I}} $\lambda\lambda$7664, 7698 \AA\ doublet in our 
medium-resolution spectra. 
However, we estimate the 2$\sigma$ upper limit for the EW of the DIB at $\lambda$ 5780 \AA\ using the same method as \citet{2013ApJ...779...38P} to be 0.24 \AA. This, together with the extinction parameter, $A_V$ = 0.58 $\pm$ 0.09 mag \citep{2020MNRAS.491.2639D}  
places \sn\ within the 1-$\sigma$ dispersion region of the relation between the EW of the 5780 \AA\ DIB and the host galaxy extinction shown in \citet[][Figure 5]{2013ApJ...779...38P}. 



\end{document}